# Feed-forward Perturbation-based Compensation for Nonlinear Optical Transmission

**Chuang Xu[1*], Alan Pak Tao Lau[1]**

[1]*Photonics Research Institute, Department of Electrical and Electronic Engineering,*

*The Hong Kong Polytechnic University, Hong Kong SAR, China*

**chuang.xu@connect.polyu.hk*

**Abstract:** We propose the feed-forward perturbation-based nonlinearity compensation method using the received signal, which outperforms conventional decision-based ones and eliminates the need for decision feedback. Additionally, combining half-half dispersion compensation with pre-/post-compensation yields further performance gains. 

## 1. Introduction

Perturbation-based (PB) analysis provides analytical approximations for fiber nonlinearity (NL) noise, facilitating the design of transmitter (Tx) or receiver (Rx) pre/post-compensation (PC) techniques to mitigate NL noise. In PB analysis, the NL noise is modeled as an additive term to the signal, and improved models such as additive-multiplicative (AM) model [1], multiplicative-additive (MA) model [2], and the second-order PB model [3] have been studied and used in the literature. Intuitively, once a PB model is chosen, the PC algorithm is readily determined as they attempt to reverse the forward model, but the correspondence is not always clear. Besides, conventionally, post-compensation requires symbol decisions for NL noise estimation, which increases complexity and can lead to errors propagation depending on the chosen algorithm. In [4], it was proposed to re-map the post-FEC bits to symbols for NL noise estimation to avoid the error symbol issue, but it will suffer from even higher complexity.

In [5], we show that the AM-model approximates the NL fiber channel well, while PC conducted in reverse MA-manner is equivalent to back propagation of the AM-model, and a raw Rx signal can directly serve as the input for the PC stage, without the need for feedback of symbol decision for NL estimation. We note that a similar idea appeared in [6]. However, the analysis in [5] is done without frequency offset (FO), laser phase noise and hence carrier frequency/phase recovery (CFR/CPR). In this paper, we compare it with the decision (Dx)-based PC in a more realistic situation, showing that although the Dx-based PC method can return a higher SNR, it can instead have a high symbol error ratio (SER) due to the feedback of decision errors, and the Rx-based feed-forward PC is superior.

## 2. Different PC Methods in Ideal and Practical Situations

Based on AM-model, given a Tx sequence $\boldsymbol{t} = [t_1, t_2, \dots, t_N]$, the Rx sequence $\boldsymbol{r} \approx (\boldsymbol{t} + \boldsymbol{B_t})e^{i\boldsymbol{A_t}} + \boldsymbol{h}$ with element-wise multiplication, $\boldsymbol{A_t}$ and $\boldsymbol{B_t}$ denote the NL phase and circularly symmetric noise, and the subscript "$\boldsymbol{t}$" indicates that they are calculated based on $\boldsymbol{t}$. We add $\boldsymbol{h}$ to include the residual NL noise that is not captured by AM-model. Note that $\boldsymbol{r}$ is a constellation rotated by angle of $\overline{\boldsymbol{A}}_{\boldsymbol{t}} = \mathbb{E}[\boldsymbol{A_t}]$ on average, when no laser phase noise exists. The proposed Rx-based PC is given by $\hat{\boldsymbol{r}}_{\boldsymbol{r}-\mathrm{MA}} = (\boldsymbol{r} - \boldsymbol{B_r})e^{-i\boldsymbol{A_r}}$, and the Dx-based PC signal is $\hat{\boldsymbol{r}}_{\boldsymbol{d}-\mathrm{AMs}} = \boldsymbol{r} - [(\boldsymbol{d} + \boldsymbol{B_d})e^{i\boldsymbol{A_d}} - \boldsymbol{d}]$, which uses the symbol decision $\boldsymbol{d}$ to estimate NL noise by AM-model and then subtract it (as indicated by "s" in the subscript). Besides, ideal PC is included for comparison and easy analysis, $\hat{\boldsymbol{r}}_{\boldsymbol{t}-\mathrm{AMs}} = \boldsymbol{r} - [(\boldsymbol{t} + \boldsymbol{B_t})e^{i\boldsymbol{A_t}} - \boldsymbol{t}]$.

First, we revisit the effectiveness of $\hat{\boldsymbol{r}}_{\boldsymbol{r}-\mathrm{MA}}$ in the ideal transmission situation with the following simulation: 16384 45 GBaud dual polarization probabilistic shaping (PS)-64QAM (shaping factor $\nu = 0.03$) symbols are generated by CCDM and LDPC code [7]. Sinc waveform (roll-off = 0) with 8 samples per symbol was used for modulation. The transmission link is 10x80km G.652 fiber ($\alpha = 0.2$ dB/km, $D = 16$ ps/km/nm, $\gamma = 1.3$ W$^{-1}$km$^{-1}$) with EDFA

(a)

| Situation | Name | Notation |
|---|---|---|
| Ideal | Rx-Rx-MA | $\hat{\boldsymbol{r}}_{\boldsymbol{r}-\mathrm{MA}}$ |
| | Rx-De/Tx-AMs | $\hat{\boldsymbol{r}}^{\varphi}_{\boldsymbol{d/t}-\mathrm{AMs}}$ |
| | Rx-De/Tx-MAs | $\hat{\boldsymbol{r}}^{\varphi}_{\boldsymbol{d/t}-\mathrm{MAs}}$ |
| Practical | Rx-Rx-MA-CPR | $\hat{\boldsymbol{r}}_{\boldsymbol{r}-\mathrm{MA-CPR}}$ |
| | CPR-CPR-MA | $\hat{\boldsymbol{c}}_{\boldsymbol{c}-\mathrm{MA}}$ |
| | CPR-De/Tx-AMs | $\hat{\boldsymbol{c}}_{\boldsymbol{d/t}-\mathrm{AMs}}$ |
| | CPR-De/Tx-MAs | $\hat{\boldsymbol{c}}_{\boldsymbol{d/t}-\mathrm{MAs}}$ |
| | CPR-De/Tx-AMacs | $\hat{\boldsymbol{c}}_{\boldsymbol{d/t}-\mathrm{AMacs}}$ |

Remarks:
1. "s" stands for subtraction;
2. Only a single CPR in included in the practical situation;
3. A final phase rotation stage (if necessary) is applied to straighten the PC constellation, which is omitted in the notation for clarity, e.g., $\hat{\boldsymbol{c}}_{\boldsymbol{c}-\mathrm{MA}}$ means $\hat{\boldsymbol{c}}_{\boldsymbol{c}-\mathrm{MA-Rot}}$.

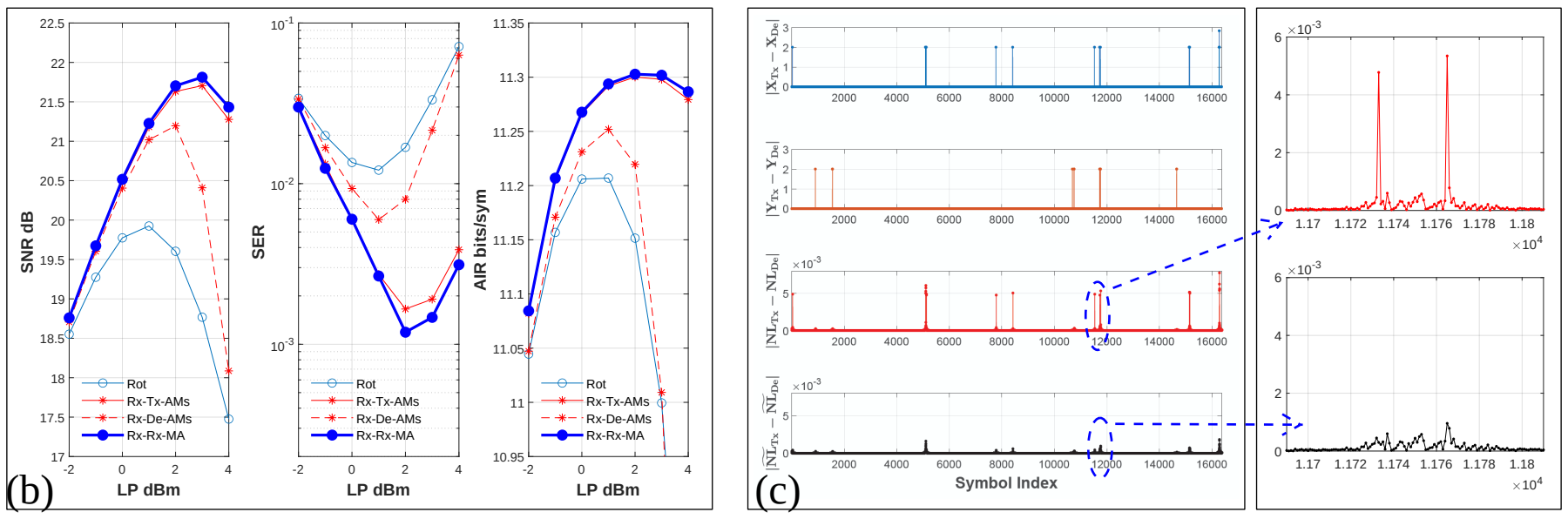

Fig. 1. (a) The notations for different PC signals; (b) Rx-Rx-MA has the same performance with Rx-Tx-AMs in the ideal transmission case; (c) Error symbol decisions induce great errors in NL noise extimation, $(\boldsymbol{d} + \boldsymbol{B_d})e^{i\boldsymbol{A_d}} - \boldsymbol{d}$, but much less errors in $(\boldsymbol{d} + \boldsymbol{B_d})e^{i\widetilde{\boldsymbol{A}}_{\boldsymbol{d}}} - \boldsymbol{d}$.

This work was supported by the Hong Kong Government Research Grants Council General Research Fund (GRF) under Project PolyU 15225423.

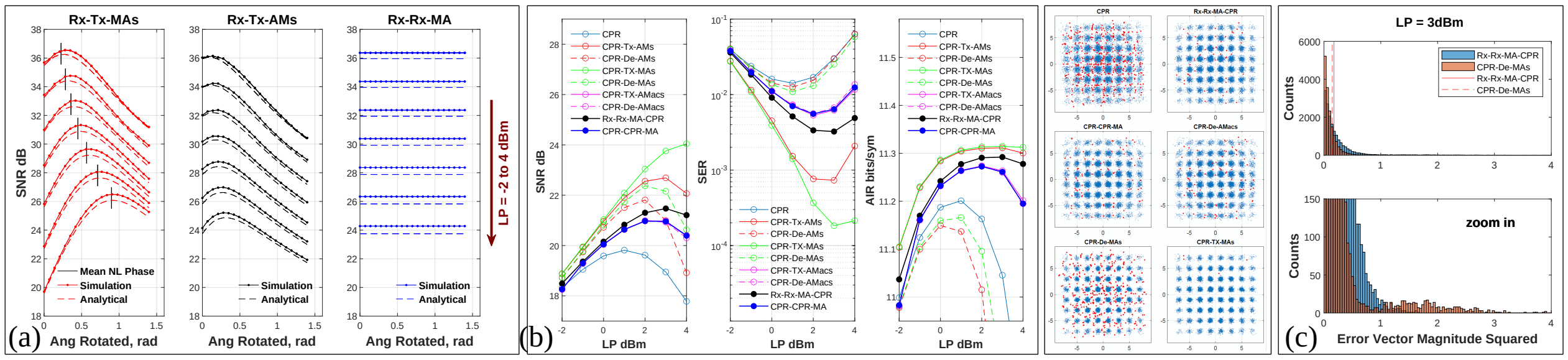

Fig. 2. (a) Simulation and analytical approximation of SNR of $\hat{\boldsymbol{r}}^{\varphi}_{\boldsymbol{t}-\mathrm{MAs}}$ and $\hat{\boldsymbol{r}}^{\varphi}_{\boldsymbol{t}-\mathrm{AMs}}$ and $\hat{\boldsymbol{r}}_{\boldsymbol{r}-\mathrm{MA}}$; (b) Performance Rx/Dx/Tx-based PC when FO and laser phase noise exist and the corresponding constellations in X-pol at LP = 3dBm; (c) Histogram of EVM$^2$ of Rx-Rx-MA-CPR and CPR-De-MAs.

amplification (NF = 5 dB). No laser phase noise, FO, nor polarization effect is included. Fig. 1(b) shows the SNR, SER and achievable information rate (AIR) curves of these three PC methods. $\hat{\boldsymbol{r}}_{\boldsymbol{r}-\mathrm{MA}}$ shows the same level of performance as $\hat{\boldsymbol{r}}_{\boldsymbol{t}-\mathrm{AMs}}$, but $\hat{\boldsymbol{r}}_{\boldsymbol{d}-\mathrm{AMs}}$ shows penalties and deteriorates sharply in the high launching power (LP) region because the high SER makes the NL estimation very inaccurate. Fig. 1(c) shows the difference between the symbol decision and Tx, and the difference between the NL noise estimation from them. It is clear that whenever a symbol decision error in X/Y polarization occurs, it incurs a significant/small error in NL estimation in X polarization, and the error spreads across multiple symbols (as shown by the zoom-in) due to dispersion-induced symbol interaction. In contrast, another De-based PC given by $\hat{\boldsymbol{r}}_{\boldsymbol{d}-\mathrm{AMacs}} = \boldsymbol{r} - \left[(\boldsymbol{d} + \boldsymbol{B_d})e^{i\widetilde{\boldsymbol{A}}_{\boldsymbol{d}}} - \boldsymbol{d}\right]$ calculates the NL noise based on the derotated AM-model constellation, by which decision errors will have much less influence on the NL noise estimation, as shown in the bottom plot of Fig. 1(c), because the estimated NL noise becomes $\widetilde{NL} \approx i\boldsymbol{d}\widetilde{\boldsymbol{A}}_{\boldsymbol{d}} + \boldsymbol{B_d}e^{i\widetilde{\boldsymbol{A}}_{\boldsymbol{d}}}$, in which the "DC" term ($\overline{\boldsymbol{A}}_{\boldsymbol{d}}$) is removed, $\widetilde{\boldsymbol{A}}_{\boldsymbol{d}} = \boldsymbol{A_d} - \overline{\boldsymbol{A}}_{\boldsymbol{d}} \approx \boldsymbol{0}$, and thus the error in $\boldsymbol{d}$ is much less propagated into NL noise estimation.

In practice, FO and laser phase noise exist and CFR/CPR are needed. Besides, the Dx-based PC has to be implemented after CPR because it needs a clearly demodulated constellation for symbol decision. Thus, there is an implicit phase derotation on $\boldsymbol{r}$, which derotates the constellation to the upright position before NL noise subtraction, which can actually increase the SNR as shown in the following. For simplicity, we analyze on Tx-based PC, and we exclude laser phase noise and ASE first, but still include the phase derotation before PC. We also add the Tx-MAs-based method, which estimates NL noise according to the MA model instead, denoted by $\hat{\boldsymbol{r}}_{\boldsymbol{t}-\mathrm{MAs}}$, because we found it shows extra gain with phase derotation. Notations of different PC methods are summarized in Fig. 1(a).

After phase derotation by $\varphi$, $\hat{\boldsymbol{r}}^{\varphi}_{\boldsymbol{t}-\mathrm{MAs}} = \boldsymbol{r}e^{-i\varphi} - \left[\boldsymbol{t}e^{i\boldsymbol{A_t}} + \boldsymbol{B_t} - \boldsymbol{t}\right] \approx \underbrace{[1 + R(\varphi)]\boldsymbol{t}}_{\text{enlarged sig}} + \underbrace{\boldsymbol{h}e^{-i\varphi} + i(\boldsymbol{A_t} - \varphi)\boldsymbol{B_t} + i\widetilde{\boldsymbol{A}}_{\boldsymbol{t}}\boldsymbol{t}R(\varphi)}_{\text{NL noise after PC}}$, $\hat{\boldsymbol{r}}^{\varphi}_{\boldsymbol{t}-\mathrm{AMs}} = \boldsymbol{r}e^{-i\varphi} - \left[(\boldsymbol{t} + \boldsymbol{B_t})e^{i\boldsymbol{A_t}} - t\right] \approx [1 + R(\varphi)]\boldsymbol{t} + \boldsymbol{h}e^{-i\varphi} + \boldsymbol{B_t}e^{i\widetilde{\boldsymbol{A}}_{\boldsymbol{t}}}R(\varphi) + i\widetilde{\boldsymbol{A}}_{\boldsymbol{t}}\boldsymbol{t}R(\varphi)$, where $R(\varphi) = e^{i(\overline{\boldsymbol{A}}_{\boldsymbol{t}} - \varphi)} - e^{i\overline{\boldsymbol{A}}_{\boldsymbol{t}}}$ captures the feature of derotation by $\varphi$, which can have an enlarging effect. Moreover, for $\hat{\boldsymbol{r}}^{\varphi}_{\boldsymbol{t}-\mathrm{MAs}}$, the NL noise after PC can be reduced when $\varphi$ approaches $\overline{\boldsymbol{A}}_{\boldsymbol{t}}$, while for $\hat{r}^{\varphi}_{\boldsymbol{t}-\mathrm{AMs}}$, thes noise power is enlarged. From simulations, we can find that $\mathrm{SNR}\left(\hat{\boldsymbol{r}}^{\varphi}_{\boldsymbol{t}-\mathrm{MAs}}\right)$ is maximized around $\overline{\boldsymbol{A}}_{\boldsymbol{t}}$, reaching $\mathrm{SNR}\left(\hat{\boldsymbol{r}}^{\varphi_{\mathrm{opt}} = \overline{\boldsymbol{A}}_t}_{\boldsymbol{t}-\mathrm{MAs}}\right) \approx \frac{|1 + R(\overline{\boldsymbol{A}}_t)|^2 P_t}{\sigma_h^2 + \sigma_{A_t}^2\sigma_{B_t}^2 + |R(\overline{\boldsymbol{A}}_t)|^2\sigma_{A_t}^2 P_t}$. In contrast, $\mathrm{SNR}(\hat{\boldsymbol{r}}^{\varphi}_{\boldsymbol{t}-\mathrm{AMs}})$ is not increased much because the NL noise is also amplified as analyzed above. Besides, we approximated $\mathrm{SNR}(\hat{\boldsymbol{r}}_{\boldsymbol{r}-\mathrm{MA}}) \approx P_t/\sigma_h^2$, which indicates that the NL distortion captured by the AM-model is fully removed and only the residual NL noise remains, and phase derotation does not affect it.

In Fig. 2(a), the theoretical prediction from above analysis and simulation result (ASE off) of $\mathrm{SNR}(\hat{\boldsymbol{r}}^{\varphi}_{\boldsymbol{t}-\mathrm{MAs}})$ are shown, which match each other quite well. Across different LPs, the peak of $\mathrm{SNR}\left(\hat{\boldsymbol{r}}^{\varphi}_{\boldsymbol{t}-\mathrm{MAs}}\right)$ is achieved when the phase derotation approaches the mean NL phase noise, $\overline{\boldsymbol{A}}_{\boldsymbol{t}}$, marked by the short black lines. Note that $\hat{\boldsymbol{r}}^{\varphi=0}_{\boldsymbol{t}-\mathrm{MAs}}$ is not a "matched" compensation way as it reconstructs NL noise according to MA model, so its performance is worse than $\hat{\boldsymbol{r}}^{\varphi=0}_{\boldsymbol{t}-\mathrm{AMs}}$ and $\hat{\boldsymbol{r}}_{\boldsymbol{r}-\mathrm{MA}}$. However, when the phase derotation is applied, its SNR can be greatly enhanced and eventually surpass them, while $\mathrm{SNR}\left(\hat{\boldsymbol{r}}^{\varphi}_{\boldsymbol{t}-\mathrm{AMs}}\right)$ only slightly increases by phase derotation. In [8], it is reported that adding a phase rotation to the pre-distortion term can increase the performance, which should be due to a similar reason.

In the following simulation of practical situation, FO of 1 GHz and Tx Rx phase noise (laser linewidth, 100 kHz), and ASE noise are included, and CFR and CPR are also added accordingly. Now, for Rx-based PC, we can either use a CD-compensated signal (***with*** FO and laser phase noise) for PC, followed by CFR and CPR, or perform CFR and CPR first, followed by PC and a phase rotation (to make it upright). These two methods are denoted by $\hat{\boldsymbol{r}}_{\boldsymbol{r}-\mathrm{MA-CPR}}$ and $\hat{\boldsymbol{c}}_{\boldsymbol{c}-\mathrm{MA}}$, respectively, and both of them include only ***one*** CPR stage for a fair performance comparison. The implicit phase derotation effect of CPR will benefit Tx-based PC as analyzed before. Fig. 2(b) shows that $\hat{\boldsymbol{c}}_{\boldsymbol{t}-\mathrm{MAs}}$ shows a significant gain in SNR over $\hat{\boldsymbol{r}}_{\boldsymbol{r}-\mathrm{MA-CPR}}$ and $\hat{\boldsymbol{c}}_{\boldsymbol{c}-\mathrm{MA}}$. The Dx-based PC also shows SNR improvement as it resembles the Tx-based one if SER is not high. It is worth noting that the peak SNR of $\hat{\boldsymbol{r}}_{\boldsymbol{r}-\mathrm{MA-CPR}}$ is higher than $\hat{\boldsymbol{c}}_{\boldsymbol{c}-\mathrm{MA}}$ (by 0.5

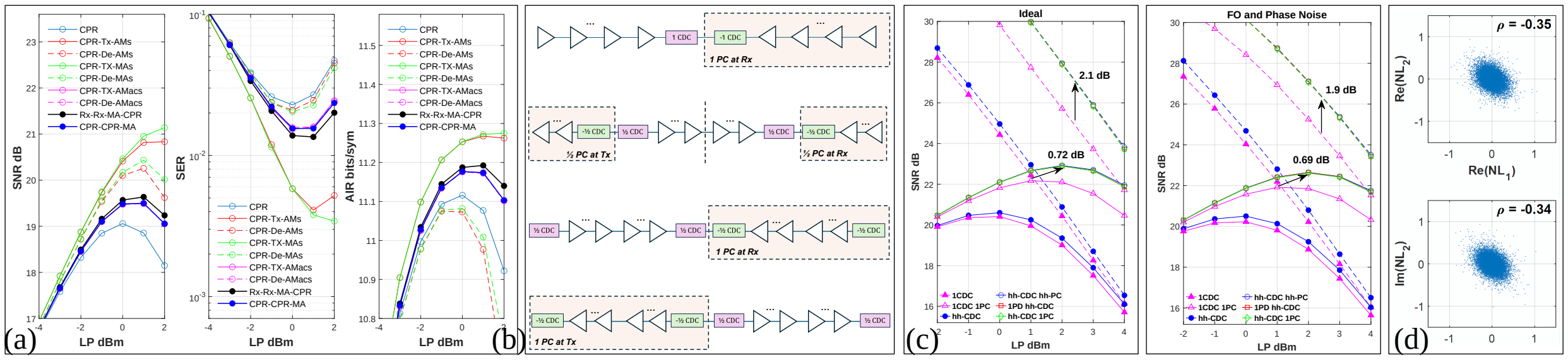

Fig. 3 (a) Performance of Rx/Dx/Tx-based PC in 7-ch WDM; (b) Link structures of full PC and hh-PC on top of full-CDC and hh-CDC; (c) SNR for structures in (b), w/ (solid) and w/o (dashed) ASE; (d) $\mathrm{Re}(\mathrm{NL}_1)$ vs $\mathrm{Re}(\mathrm{NL}_2)$, $\mathrm{Im}(\mathrm{NL}_1)$ vs $\mathrm{Im}(\mathrm{NL}_2)$, which are negatively correlated.

dB) which we attribute to the fact that CPR removes a certain amount of long-correlated NL phase noise [10], which harms the "back propagation" process due to loss of NL phase noise information.

Thus, the SNR of Dx-based PC signal can surpass the proposed Rx-based PC when CPR exists. However, we found that the corresponding SER and AIR are actually deteriorated compared with $\hat{\boldsymbol{r}}_{\boldsymbol{r}-\mathrm{MA-CPR}}$, as shown in Fig. 2(b). The corresponding constellations provide a straightforward illustration, in which the error symbols are marked with red color. Overall, both $\hat{\boldsymbol{c}}_{\boldsymbol{t}-\mathrm{MAs}}$ and $\hat{\boldsymbol{c}}_{\boldsymbol{d}-\mathrm{MAs}}$ have tight constellation clusters (thus the high SNR), but the latter has outliers that are caused by decision errors, which do not reduce the SNR much as their quantity is relatively small, but they lie across the decision boundary and thus strongly increase SER. This can also be seen from the histogram for squared error vector magnitude ($\mathrm{EVM}^2$) of $\hat{\boldsymbol{r}}_{\boldsymbol{r}-\mathrm{MA-CPR}}$ and $\hat{\boldsymbol{c}}_{\boldsymbol{d}-\mathrm{MAs}}$ in Fig. 2(c). The $\mathrm{EVM}^2$ of $\hat{\boldsymbol{c}}_{\boldsymbol{d}-\mathrm{MAs}}$ is more concentrated at 0 but has a long tail, although it has a lower average value of $\mathrm{EVM}^2$. In contrast, $\hat{\boldsymbol{c}}_{\boldsymbol{d}-\mathrm{AMacs}}$ and demonstrate close performance to $\hat{\boldsymbol{c}}_{\boldsymbol{t}-\mathrm{AMacs}}$, due to the its insusceptibility to decision errors. Moreover, both of them perform very closely to $\hat{\boldsymbol{c}}_{\boldsymbol{c}-\mathrm{MA}}$. This is not surprising because for PC implemented on signal after CPR, the actual NL phase noise has been partially removed by CPR, but the estimated NL noise is "full", so they will not effectively cancel each other out. This is similar to the situation of $\hat{\boldsymbol{c}}_{\boldsymbol{c}-\mathrm{MA}}$—they both lose part of the NL phase noise due to implementation after CPR, while they eliminate the NL noise by "estimation-subtraction" and "back propagation", respectively.

We further investigated in N-channel (N = 3, 5, 7) WDM simulations with FO and Tx Rx phase noise, the center channel is filtered and demodulated, followed by Rx/Dx/Tx-based PC. Similar to the single-channel case, the phase derotation by CPR increases $\mathrm{SNR}(\hat{\boldsymbol{r}}^{\varphi}_{\boldsymbol{d}-\mathrm{MAs}})$, with the optimal derotation phase around the mean NL phase, which is N-folds that of single-channel, but the corresponding SER and AIR are not improved much. Again, $\hat{\boldsymbol{c}}_{\boldsymbol{t}-\mathrm{AMacs}}$, $\hat{\boldsymbol{c}}_{\boldsymbol{d}-\mathrm{AMacs}}$ and $\hat{\boldsymbol{c}}_{\boldsymbol{c}-\mathrm{MA}}$ return almost the same performance, and $\hat{\boldsymbol{r}}_{\boldsymbol{r}-\mathrm{MA-CPR}}$ performs the best, with a 0.16 dB SNR gain over $\hat{\boldsymbol{c}}_{\boldsymbol{c}-\mathrm{MA}}$ in 7-ch WDM.

## 3. Half-half pre- and post-compensation at Tx and Rx

It was proposed that using half pre-CD compensation (CDC) to form a symmetry dispersion map,i.e., half-half (hh)-CDC, can effectively reduce the complexity for PC, while the PC is fully done at Tx [9]. Based on this, we split the "full PC" into a half-half pre-/post-compensation ("hh-PC") at Tx/Rx, by which the whole link is split into two sub-links so each of them has inherent symmetry (Fig.3(b)). In single-channel ideal transmission of 10x80 km with 30 GBaud PS-64QAM, we found that "hh-CDC" itself only slightly increases the SNR, but "hh-PC" on top of it reduces NL noise by 2.1 dB over full CDC+full PC, which translates into a 0.72 dB gain in peak-SNR with ASE loaded. With phase noise and FO, the gains slightly reduce to 1.9 and 0.69 dB, respectively.

As the overall NL noise can be approximated (to the first order) by the sum of the NL noise from the first and second sub-links, we extract the NL noise from each sub-link and find they are negatively correlated, with correlation coefficients up to about -0.3 to -0.4, making the overall NL noise power less than the sum of them. We note that the correlation strength depends on system parameters such as baudrate, link configuration, and constellation hierarchy, etc., and need further investigation. Roughly, the correlation strength tends to be stronger when the NL noise exhibits stronger phase-noise pattern, which becomes weaker when the span number increases [10]. However, the results also indicate that, with hh-CDC implemented, the performance of full pre-PC, full post-PC, and hh-PC is nearly the same.

## 4. Conclusion

We analyze various perturbation-based post-compensation methods, and demonstrate that Rx-based nonlinearity post-compensation outperforms the conventional Dx-based counterpart by achieving higher performance and eliminating the need for decision feedback. Furthermore, applying pre- or post-compensation in addition to half-half dispersion compensation provides additional performance gains.